\documentclass[twocolumn,pre,showpacs]{revtex4-1}
\usepackage{graphicx}
\usepackage{amsmath}
\usepackage{amsfonts}
\usepackage{amssymb}
\usepackage{float}
\usepackage{mathrsfs}

\newcommand{\omegabold}{\boldsymbol\omega}

\begin{document}
\title{Stability of tangential discontinuity for the vortex pancakes}

\author{D.\,S.\,Agafontsev$^{(a),(b)}$, E.\,A.\,Kuznetsov$^{(b),(c),(d)}$, A.\,A.\,Mailybaev$^{(e)}$}

\affiliation{\textit{
$^{(a)}$ P.P. Shirshov Institute of Oceanology of RAS, 117997 Moscow, Russia\\
$^{(b)}$ Skolkovo Institute of Science and Technology, 121205 Moscow, Russia\\
$^{(c)}$ P.N. Lebedev Physical Institute of RAS, 119991 Moscow, Russia\\
$^{(d)}$ L.D. Landau Institute for Theoretical Physics of RAS, 142432 Chernogolovka, Russia\\
$^{(e)}$ Instituto de Matem\'atica Pura e Aplicada -- IMPA, 22460-320 Rio de Janeiro, Brazil
}}
\email{dmitrij@itp.ac.ru}

\begin{abstract}
Within the incompressible three-dimensional Euler equations, we study the pancake-like high vorticity regions, which arise during the onset of developed hydrodynamic turbulence. 
We show that these regions have an internal fine structure consisting of three vortex layers. 
Such a layered structure, together with the power law of self-similar evolution of the pancake, prevents development of the Kelvin-Helmholtz instability.
\end{abstract}

\maketitle

%----------------------------------------------------------------------------
%----------------------------------------------------------------------------

\textbf{1.} 
According to the classical works of Kolmogorov and Obukhov~\cite{kolmogorov1941local,obukhov1941spectral}, in the regime of developed hydrodynamic turbulence at large Reynolds numbers $\mathrm{Re}\gg 1$, the vorticity fluctuations have power-law scaling $\langle\delta\omega\rangle\propto\varepsilon^{1/3}l^{-2/3}$ at scale $l$ from the inertial interval, i.e. they diverge at $l\to 0$; here $\varepsilon$ is the energy dissipation rate per unit volume.
Meanwhile, the time $T$ of energy transfer from large to small scales is finite and depends only on $\varepsilon$ and the integral (large) scale $L$, $T\propto {L^{2/3}}{\varepsilon^{-1/3}}$. 
Thus, the Kolmogorov-Obukhov theory points to the possibility of collapse when vorticity tends to infinity at some point in space in a finite time, and at the scales of the inertial interval this process can be studied using Euler's hydrodynamics. 
Apparently, this question was first raised in the classical work of L. Onsager in 1949~\cite{onsager1949statistical}; see article~\cite{eyink2006onsager} and book~\cite{frisch1999turbulence} for the history of this issue.
Nowadays, the existence of collapse (also called blowup) is one of the central problems in the theory of developed hydrodynamic turbulence.
One of the most discussed types of blowup is related to the compressing vortex sheets (pancakes)~\cite{saffman1981dynamics}, which were first observed in the numerical experiments~\cite{brachet1992numerical}.

On a vortex pancake, the tangential velocity component experiences a jump $\Delta V$ at a small pancake thickness $\ell_{1}$.
If we assume that this jump remains finite, and the pancake thickness tends to zero, then such a flow transforms into a velocity discontinuity subjected to the Kelvin-Helmholtz (KH) instability~\cite{landau2013fluid}.
The growth rate of this instability $\gamma$ increases linearly with the wavenumber $k$ along the discontinuity, $\gamma = k\Delta V/2$. 
If the thickness $\ell_{1}$ is finite, then the instability should saturate at large $k$, so that the maximum increment can be estimated as $\gamma_{\max} \sim \Delta V/\ell_{1}$. 
Note that vortex pancakes have been recently discussed in the context of developed hydrodynamic turbulence~\cite{migdal2021vortex}.

In a series of numerical experiments~\cite{agafontsev2015development,agafontsev2016development,agafontsev2016asymptotic}, we have studied evolution of the high vorticity regions within the framework of the incompressible three-dimensional Euler equations. 
We have confirmed that these regions represent compressing pancake-like structures and found that the flow near them is described locally by a new exact self-similar solution of the Euler equations, which combines a shear flow with an asymmetric straining flow. 
Note that, in Euler's hydrodynamics, the pancake-like structures were observed for the first time in the numerical experiments by M. Brachet et al.~\cite{brachet1992numerical}. 
In contrast to this work, we have demonstrated that the exponential growth of the vorticity maximum $ \omega_{\max} = \max | \omegabold | \propto e^{\beta_{2} t} $ and the exponential decrease of the pancake thickness $ \ell_{ 1} \propto e^ {- \beta_{1} t} $ are characterized by significantly different exponents, $ \beta_{2} / \beta_{1} \approx 2/3 $, leading to the Kolmogorov-type scaling law
\begin{equation}\label{2/3}
	\omega_{\max}\propto \ell_{1}^{-2/3}.
\end{equation}
This law is confirmed numerically for most of the pancakes; also, in~\cite{agafontsev2018development}, we have presented analytical arguments in favor of the $2/3$-scaling~(\ref{2/3}) by using the vortex line representation~\cite{kuznetsov1998hamiltonian}. 
The number of pancakes increases with time, and they provide the leading contribution to the energy spectrum.
In particular, for some initial conditions \cite{agafontsev2016development,agafontsev2019statistical}, we have observed formation of the Kolmogorov spectrum $ E_{k} \propto k^{- 5/3} $ and the power-law scalings for the longitudinal and transverse structure functions of the velocity, in a fully inviscid flow.

Taking into account the exponential growth of the vorticity maximum and the exponential decrease of the pancake thickness, the maximum increment of the KH instability for the pancake structure should be characterized by the exponential time dependency,
\begin{equation}\label{KHmax}
	\gamma_{max} \sim \Delta V/\ell_{1} \simeq \omega_{\max} \propto e^{\beta_{2}t},
\end{equation}
which would seem to indicate a double exponential amplification of the perturbation.
However, in numerical experiments~\cite{agafontsev2015development, agafontsev2016development, agafontsev2016asymptotic, agafontsev2018development, agafontsev2019statistical} we have not observed this type of instability.

In this paper, we provide several arguments to explain this fact. 
In particular, we show that the KH instability is suppressed by the self-similar shear flow of the pancake. 
Additionally, the pancakes have a fine internal structure consisting of three vortex layers, which may also prevent development of the KH instability.\\

%----------------------------------------------------------------------------
%----------------------------------------------------------------------------

\textbf{2.} We consider evolution of the high vorticity regions in the framework of the incompressible 3D Euler equations, which in the dimensionless form can be written as 
\begin{equation}\label{Euler}
	\frac{\partial\mathbf{v}}{\partial t} + (\mathbf{v}\cdot\nabla)\mathbf{v} = -\nabla p, \quad\quad \mathrm{div}\,\mathbf{v} = 0.
\end{equation}
Here $ \mathbf{v} $ is the velocity field and $ p $ is the pressure. 
As has been shown in~\cite{agafontsev2016asymptotic}, the flow near the pancake is described by the following exact solution of the Euler equations,
\begin{eqnarray}
	\mathbf{v}(\mathbf{x},t) &=& -\omega_{\max}\ell_{1}\,f\left(\frac{x_1}{\ell_{1}}\right) \mathbf{n}_{3} + 
	\left(\begin{array}{c}
		-\beta_{1}x_{1} \\ \beta_{2}x_{2} \\ 
		\beta_{3}x_{3}\end{array}\right),  \label{solution-1} \\
	p &=& - \frac{\beta_{1}x_{1}^{2}}{2} - \frac{\beta_{2}x_{2}^{2}}{2} - \frac{\beta_{3}x_{3}^{2}}{2}, \label{solution-1p} \\
	\omegabold(\mathbf{x},t) &=& \omega_{\max}f^{\prime}\left(\frac{x_1}{\ell_{1}}\right) \mathbf{n}_{2}.  \label{solution-2}
\end{eqnarray} 
Here $ \omegabold = \mathbf{rot} \, \mathbf{v} $ is the vorticity field, $ \beta_{1} $, $ \beta_{2} $ and $ \beta_{3} $ are arbitrary constants such that $ - \beta_{1} + \beta_{2} + \beta_{3} = 0 $, time-dependent functions $ \omega_{\max} = \omega_{0} e^{\beta_{2} t} $ and $ \ell_{1} = h_{0} e^{- \beta_{1} t} $ describe evolution of the vorticity maximum and the pancake thickness, and $ f (\xi) $ is an arbitrary smooth function with first derivative attaining the maximum at zero, $ \max f^{\prime} (\xi) = f^{\prime} (0) = 1 $.
Solution~(\ref{solution-1})-(\ref{solution-1p}) is written in the Cartesian coordinate system, whose origin is taken at the vorticity maximum, the $x_{1}$-axis is perpendicular to the pancake and the $x_{2}$-axis is directed along the vorticity vector; the solution has infinite energy and describes the pancake structures developing in the numerical experiments only locally.

Solution~(\ref{solution-1})-(\ref{solution-1p}) does not impose restrictions on the ratio of exponents $ \beta_{2} / \beta_{1} $, which defines the power law scaling $ \omega_{\max} \propto \ell_{1}^{- \beta_{2} / \beta_{1}} $. 
However, as we have shown in~\cite{agafontsev2015development, agafontsev2016development}, most of the pancakes follow the $2/3$-scaling~(\ref{2/3}) corresponding to the ratio $ \beta_{2} / \beta_{1} = 2/3 $; see also~\cite{agafontsev2018development}.\\

%----------------------------------------------------------------------------
%----------------------------------------------------------------------------

\textbf{3.} Let us analyze the possible instability of the pancake-like regions of high vorticity. 
First, we note that the problem of their stability has not been solved yet, since the solution~(\ref{solution-1})-(\ref{solution-1p}) differs significantly from the stationary velocity discontinuity. 
In particular, the pancake structure is characterized by the stationary straining and compressing shear flows; also, it is highly degenerate~\cite{agafontsev2018development}.
These features may stabilize the flow. 

It is also important that, according to Eq.~(\ref{solution-1}), the jump in the tangential velocity at the pancake thickness is proportional to $ \omega_{\max} \ell_{1} $, and in the case of $ \beta_{1}> \beta_{ 2} $ (in particular, this is true for the $2/3$-scaling~(\ref{2/3}) when $ \beta_{2} = 2 \beta_{1} / 3> 0 $) this product decreases exponentially, $ \omega_{ \max} \ell_{1} \to 0 $. 
\textit{This means that, at long time, the solution~(\ref{solution-1})-(\ref{solution-1p}) does not transform to a velocity discontinuity. 
Instead, the jump in the tangential velocity tends to zero}.

Second, in the numerical experiments~\cite{agafontsev2015development, agafontsev2016development, agafontsev2016asymptotic, agafontsev2018development, agafontsev2019statistical} we have not seen any characteristic signs of the KH instability. 
In the simulations, we have used the pseudospectral Runge-Kutta fourth-order method, in which all the spatial derivatives are calculated via the fast Fourier transform, and the adaptive grid, which resolves optimally the perpendicular direction of the main vorticity pancake and is adjusted automatically based on the analysis of the Fourier spectrum of the solution~\cite{agafontsev2015development}. 
With these methods, the energy $ E = (1/2) \int \mathbf{v}^{2} \, d \mathbf{x}^{3} $ and the helicity $ \Omega = \int (\omegabold \cdot \mathbf{v}) \, d \mathbf{x}^{3} $ are conserved with a relative error of $ 10^{- 11} $, and the numerical experiments carried out in significantly different grids converge ideally with each other~\cite{agafontsev2016development}.
Additionally, the numerical experiments that use the direct numerical integration of the Euler equations agree very well with the experiments in the vortex line representation~\cite{agafontsev2018development}.
The latter is important from the point of view of the accuracy and control of numerical simulations, since the vortex line representation is the result of partial integration of the Euler equations, in which the Cauchy invariants are preserved along each vortex line trajectory~\cite{kuznetsov1998hamiltonian}.

The most accurate experiments have been performed in~\cite{agafontsev2016asymptotic,agafontsev2018development} for two different initial flows $ I_{1} $ and $ I_{2} $ in grids with the total number of nodes $2048^3$. 
At the final time of the experiments, the global vorticity maximum demonstrated an increase by $ 12.2 $ and $ 7.6 $ times, respectively, the densest axis of the grid contained more than $4000$ points, and the thickness of the main pancakes, determined as full width at half maximum of the vorticity, corresponded to only $10$-$12$ nodes of the grid. 
As shown in~\cite{agafontsev2016asymptotic}, during the evolution, the main pancake structure remains parallel to itself and the roll-up phenomenon characteristic to the KH instability is absent.

\begin{figure}[t]
	\centering
	\includegraphics[width=9cm]{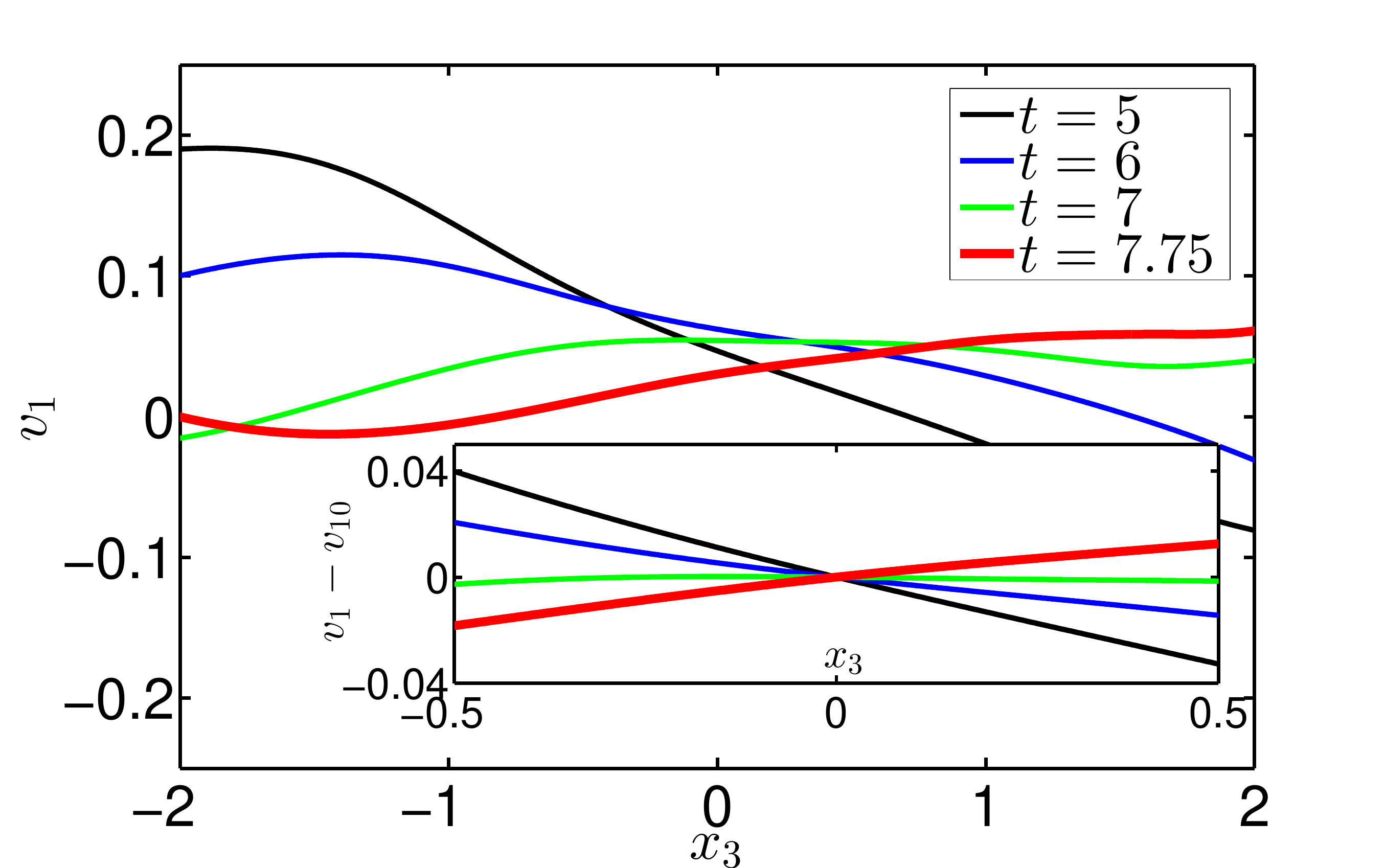}
	
	\caption{{\it (Color on-line)} 
	The velocity component $v_{1}$ along axis $x_{3}$ at different times, in the pancake's coordinates corresponding to solution~(\ref{solution-1})-(\ref{solution-1p}). 
	The inset shows zoom of the dependency $(v_{1}-v_{10})$ vs. $x_{3}$; here $v_{10}$ is the first component of the velocity $v_{1}$ at $x_{3}=0$. 
	Data corresponds to the main pancake structure developing in the numerical experiment~\cite{agafontsev2016asymptotic}.
	}
	\label{fig:fig1}
\end{figure}

In terms of solution~(\ref{solution-1})-(\ref{solution-1p}), development of the KH instability should lead to the appearance of sharp gradients of the velocity component $ v_{1} $, perpendicular to the pancake, along the pancake longitudinal direction $ x_{3} $.
However, as shown in Fig.~\ref{fig:fig1} for the main vortex pancake developing from the initial flow $ I_{1} $, the component $ v_{1} $ is close to zero and practically does not change along the $x_{3}$-axis even at the end of the numerical experiment. 
The presence of sharp gradients of $v_{1}$ along $x_{3}$-axis would mean excitation of high harmonics in the velocity field that correspond to the direction $k_{3}$.
Then, due to the relation $ | \omegabold (\mathbf{k}) | = | \mathbf{k} | \cdot | \mathbf{v} (\mathbf{k}) | $ between the Fourier-transformed velocity $ \mathbf{v} (\mathbf{k}) $ and vorticity $ \omegabold (\mathbf{k}) $, valid for an incompressible flow, the vorticity should also contain high harmonics along the same $ k_{3} $ direction. 
However, in~\cite{agafontsev2015development} we have shown that the vorticity field contains high harmonics corresponding to the perpendicular direction to the pancake $k_{1}$ only.

\begin{figure}[t]
	\centering
	\includegraphics[width=9cm]{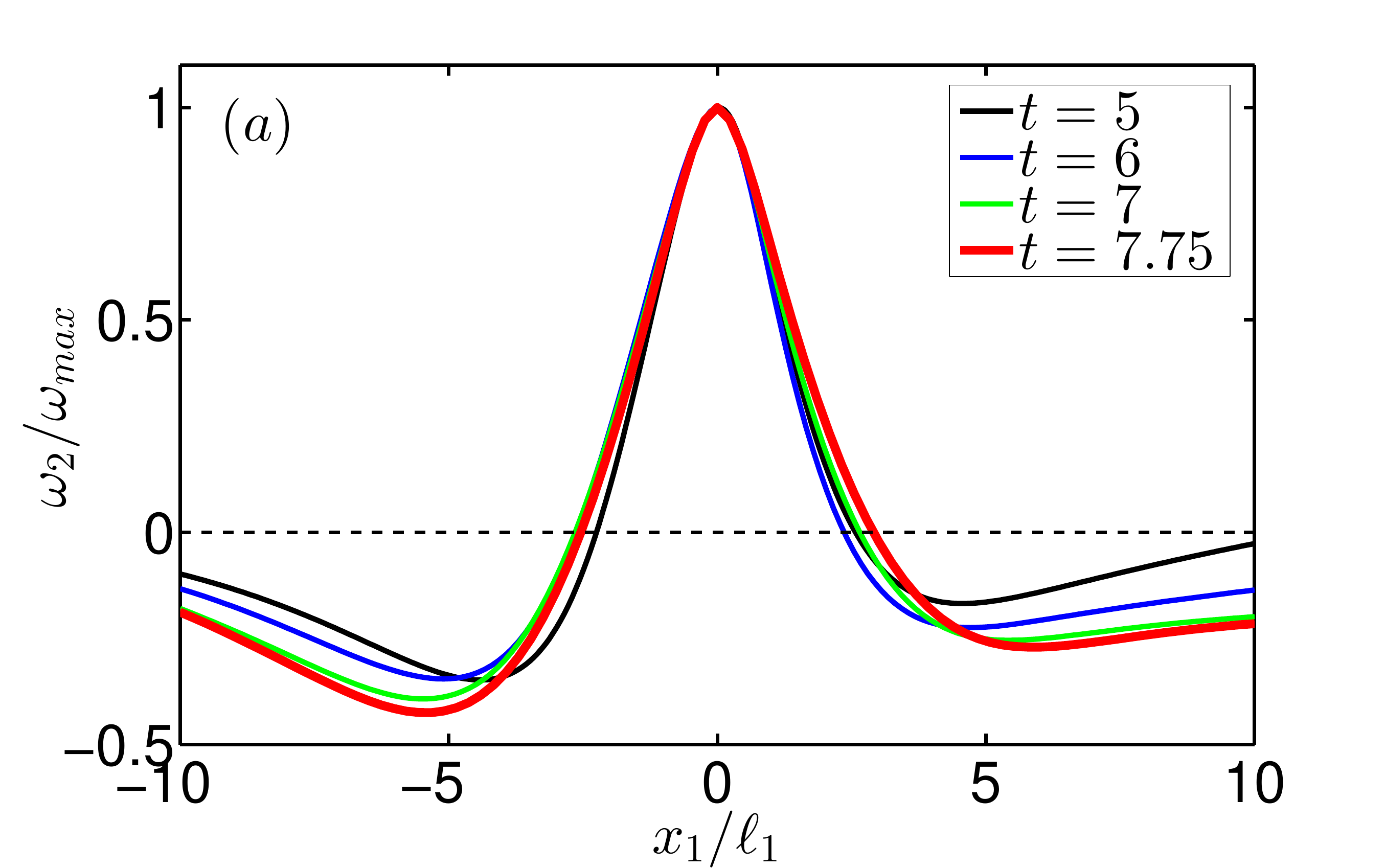}\\
	\includegraphics[width=9cm]{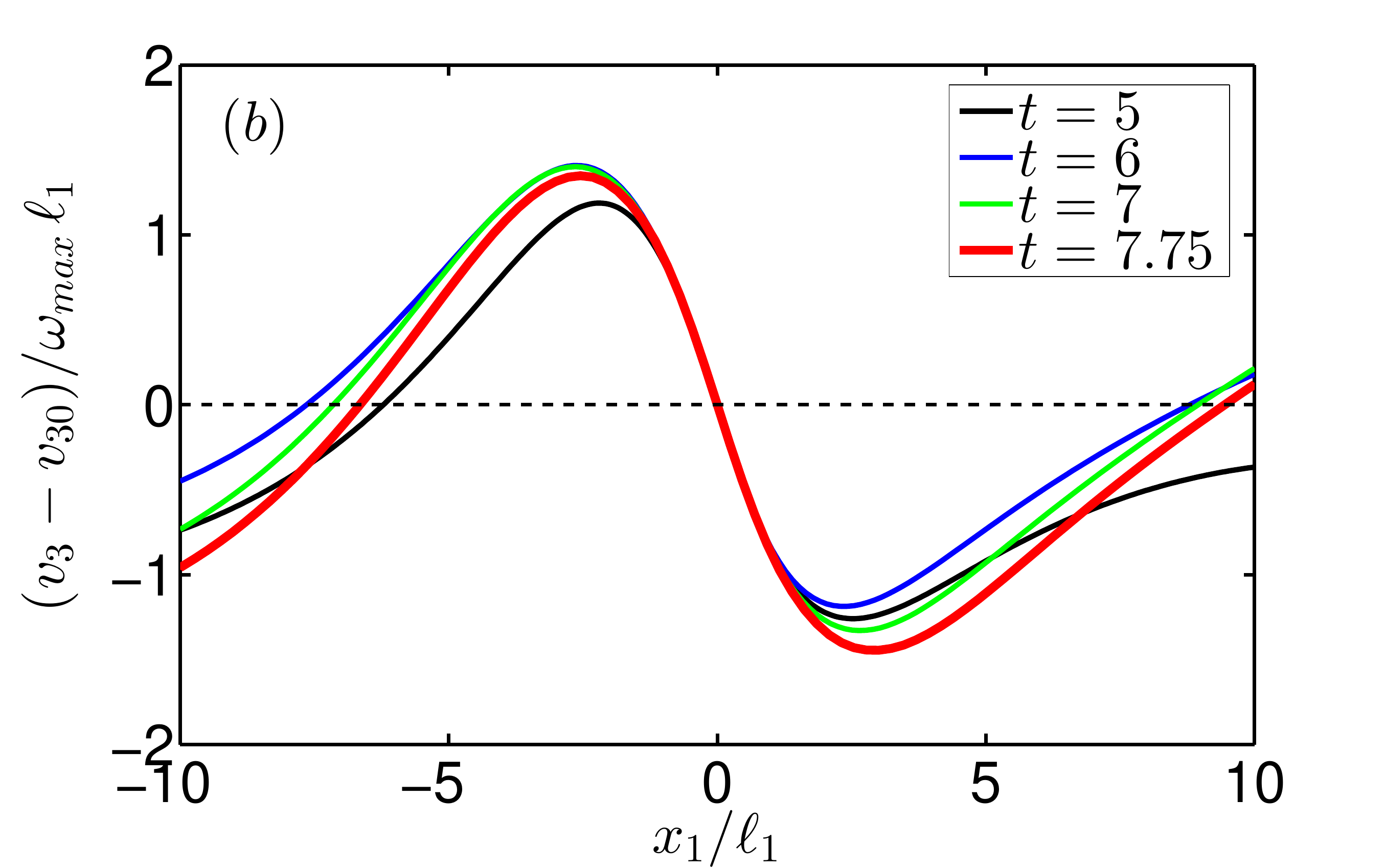}
	
	\caption{{\it (Color on-line)} 
	(a) Normalized second component of the vorticity $\omega_{2}/\omega_{\max}$ and (b) normalized third component of the velocity $(v_{3}-v_{30})/\omega_{\max}\ell_{1}$ as functions of $\xi = x_{1}/\ell_{1}$ at different times, in the pancake's coordinates corresponding to solution~(\ref{solution-1})-(\ref{solution-1p}); here $v_{30}$ is the third component of the velocity $v_{3}$ at $x_{1}=0$. 
	The black dashed lines indicate zero. 
	Data corresponds to the main pancake structure developing in the numerical experiment~\cite{agafontsev2016asymptotic}.
	}
	\label{fig:fig2}
\end{figure}

The similar results have been obtained for other pancake structures developing from the initial flow $ I_{1} $. 
Numerical experiments performed in~\cite{agafontsev2018development} for the initial flow $ I_{2} $, as well as in~\cite{agafontsev2016development, agafontsev2019statistical} for several dozens of random initial flows in grids limited by $ 1024^3 $ nodes, have not revealed any signs of the KH instability as well.

Third, in case of the instability, the gain coefficient $ \Theta (k, t) $ for the perturbation harmonic $ k $ over the time $ t $ should behave as
\begin{equation}
	\log\Theta(k,t) \propto k\int_{0}^{t}\Delta V(\tau) \,d\tau. \label{Gamma(k,t)}
\end{equation}
Here we assume that the instability increment is proportional to the jump in the tangential velocity and the wave number, $ \gamma \propto k \Delta V $.
Taking into account the $2/3$-scaling~(\ref{2/3}), the velocity jump exponentially decreases with time, $ \Delta V \propto \omega_{\max} \ell_{1} \propto \ell_{1}^{1/3} \propto e^{- \beta_{1} t / 3} \to 0 $.
Thus, harmonics of the perturbation grow slower than exponentially, and at long time the gain coefficient becomes saturated, 
\begin{equation}
	\lim_{t\to +\infty}\bigg[\log\Theta(k,t)\bigg] \propto \frac{k}{\beta_{1}}. \label{Gamma(k,t)_asymptotic}
\end{equation}
Consequently, if high harmonics are not excited in the initial flow (that corresponds to the onset of developed hydrodynamic turbulence), then they will remain small due to the finiteness of the gain coefficient~(\ref{Gamma(k,t)_asymptotic}). 
Effectively, this means absence of the instability.

Note that, in the relation~(\ref{KHmax}), the exponentially increasing with time maximum increment of the instability acts on harmonic with ever-increasing wavenumber $ k \sim 1 / \ell_{1} $, i.e., this relation does not describe dynamics of any given perturbation harmonic. 
On the contrary, we may expect that the straining components of the velocity field acting along the pancake plane, see Eq.~(\ref{solution-1}), should lead to decrease of the wavenumber for any perturbation with time.

Fourth, the fine structure of the pancake, defined in the model~(\ref{solution-1})-(\ref{solution-1p}) by the function $ f (\xi) $ and determining the jump in the tangential velocity component,
\begin{equation}
	\Delta V = \omega_{\max}\ell_{1}[f(-a)-f(a)], \label{DeltaV}
\end{equation}
between the points $ x_{1} = \pm a \ell_{1} $, may also prevent development of the KH instability. 
Indeed, the straining velocity field $ (- \beta_{1} x_{1}, \beta_{2} x_{2}, \beta_{3} x_{3})^{T} $ is antisymmetric with respect to mirror transformation $ \mathbf{x} \to - \mathbf{x} $. 
Therefore, it is natural to assume that the function $ f (\xi) $ may also be antisymmetric. 
Note that, in the solution~(\ref{solution-1})-(\ref{solution-1p}), one can always add an arbitrarily time-dependent velocity $ (0, V_{2} (t), V_{3} (t ))^{T} $, which will only lead to the appearance of additional terms in the pressure; hence, we can choose $ f (0) = 0 $.
In the simplest case of antisymmetry of $ f (\xi) $, and under the assumption that this function is localized, $ \lim_{\xi \to \pm \infty} f (\xi) = 0 $, its first derivative $ f^{\prime} (\xi) $ is symmetric with the maximum at $\xi = 0$ and has two negative ``satellites''. 
Very similar dependencies for the functions $ f^{\prime} (\xi) $ and $ f (\xi) $ are observed in the numerical experiments, see for instance Fig.~\ref {fig:fig2} for the main pancake developing from the initial flow $ I_{1} $.
The figure shows at different times the dependencies on $ x_{1} / \ell_{1} $ for the normalized second vorticity component $ \omega_{2} / \omega_{\max} $ [corresponds to the function $ f^{\prime} (\xi) $] and the normalized third velocity component $ (v_{3} -v_{30}) / \omega_{\max} \ell_{1} $ [corresponds to $ -f (\xi) $]; here $ v_{30} $ is the value of the third velocity component $ v_{3} $ at $ x_{1} = 0 $.
Similar dependencies are observed for other pancakes and other initial flows.

Note that, as follows from Fig. \ref{fig:fig2}, at distances of about ten pancake thicknesses $ x_{1} / \ell_{1} \sim 10 $, the self-similar evolution of the flow is not established yet.
Therefore, the normalized velocity and vorticity at such distances are poorly related to the functions $ f (\xi) $ and $ f^{\prime} (\xi) $.
We believe that this behavior is connected to the fact that, as shown in~\cite{agafontsev2015development, agafontsev2016asymptotic}, the pancake regions are slightly curved, deviating from the pure plane by much larger than the pancake thickness. 
This leads to interaction of different pancake segments through the ``intermediate'' flow, that may prevent establishment of the self-similar regime at the intermediate distances between the self-similar pancake at $ x_{1} \sim \ell_{1} $ and the non-self-similar region at $ | x_1 | \gg \ell_{1} $.

Thus, the pancakes have a fine layered structure that consists of a pronounced central maximum of vorticity and the two practically symmetric satellites with oppositely directed vorticity. 
As noted above, such a layered structure may mean localization of the function $ f (\xi) $, which in turn should suppress the velocity jump~(\ref{DeltaV}) measured at many pancake thicknesses $ a \gg 1 $, preventing development of the KH instability.\\

%----------------------------------------------------------------------------
%----------------------------------------------------------------------------

\textbf{4.} In this paper, we have considered the possibility of the Kelvin-Helmholtz instability for the pancake-like high vorticity regions, developing in the framework of the incompressible three-dimensional Euler equations. 
We have shown that this instability does not appear in the numerical experiments, that may be associated both with the self-similar evolution of the shear flow of the pancake and with the presence of internal pancake structure. 
In particular, the jump in the tangential velocity component decreases with time as cubic root of the pancake thickness, $ \Delta V \propto \ell_{1}^{1/3} \to 0 $, and therefore, at long time, the pancake does not transform into the tangential discontinuity. 
Also, the pancakes have a fine layered structure that consists of a pronounced central maximum of vorticity and two practically symmetric satellites with oppositely directed vorticity. 
The presence of such a layered structure may indicate to the localization of the function $ f (\xi) $ in the solution~(\ref{solution-1})-(\ref{solution-1p}), that should mean absence of a significant jump in the tangential velocity and, hence, suppression of the Kelvin-Helmholtz instability.\\

\textit{Acknowledgments}. 
The work of D.\,S. Agafontsev and E.\,A. Kuznetsov was supported by the Russian Science Foundation (grant 19-72-30028). 
The simulations were performed at the Novosibirsk Supercomputer Center (NSU), while the analysis of the results was done at the Data Center of IMPA (Rio de Janeiro). 
D.\,S. Agafontsev acknowledges the support from IMPA during the visits to Brazil. 
A.\,A. Mailybaev is supported by CNPq grants 303047/2018-6, 406431/2018-3. 

%----------------------------------------------------------------------------
%----------------------------------------------------------------------------

\end{document}